\documentclass[twocolumn]{aa}
\usepackage{graphicx}
\usepackage{lscape}
\usepackage[figuresright]{rotating}
\usepackage{txfonts}
\begin{document}

   \title{Looking at quasars through galaxies}

   \author{L. \"{O}stman\inst{1}
          \and
	  A. Goobar\inst{2}
	  \and
          E. M\"{o}rtsell\inst{3}
	  \fnmsep
          }

   \offprints{L. \"{O}stman}

   \institute{Department of Physics, Stockholm University, SE 106 91 Stockholm, Sweden \\
              \email{linda@physto.se}
         \and
   	     Department of Physics, Stockholm University, SE 106 91 Stockholm,
	     Sweden \\
             \email{ariel@physto.se}
	 \and
             Department of Astronomy, Stockholm University, 
	     SE 106 91 Stockholm, Sweden \\
             \email{edvard@astro.su.se}
             }

   \date{Received -; accepted -}

   \abstract{

Observations of quasars (QSOs) shining through or close to galaxies
offer a way to probe the properties of the foreground matter through
dust extinction and gravitational lensing. In this paper the
feasibility of measuring the dust extinction properties is
investigated using the backlitting of QSOs. We test our method to
search for QSOs affected by intervening extinction, by matching the
coordinates in the SDSS QSO DR3 catalogue with the New York University
Value-Added Galaxy Catalog. In total, 164 QSO-galaxy pairs were found
with a distance of less than 30 kpc between the galaxy centre and the
QSO line-of-sight at the galaxy redshift. Investigating the QSO
colours with multiband SDSS photometry, two pairs with galaxy
redshifts $z < 0.08$ were found to be particularly interesting in that
the QSOs show evidence of heavy Galactic type extinction with $R_V\sim
3.1$ at very large optical radii in the foreground spiral
galaxies. With the available data, it remains inconclusive whether the
two pairs can be explained as statistical colour outliers, by host
extinction or if they provide evidence of dust in the outskirts of
spiral galaxies. Deeper galaxy catalogues and/or higher resolution
follow-up QSO spectra would help in resolving this problem. We also
analyse five QSOs reported in the literature with spectroscopic
absorption features originating from an intervening system. These
systems are at higher redshifts than the other two and we find in most
cases significantly lower best fit values of $R_V$. The wide range of
preferred values of $R_V$ found, although affected by substantial
uncertainties, already indicates that the dust properties in other
galaxies may be different from the Milky Way. Furthermore, the
available data suggests a possible evolution in the dust properties
with redshift, with lower $R_V$ at high $z$.
\keywords{dust,extinction -- galaxies: ISM -- galaxies: evolution --
quasars: general -- techniques: photometric } }

\maketitle


\section{Introduction}
\label{sec:intro}

\begin{figure}
  \centering \includegraphics[bb= 110 100 280 280, clip=true,
  width=6cm]{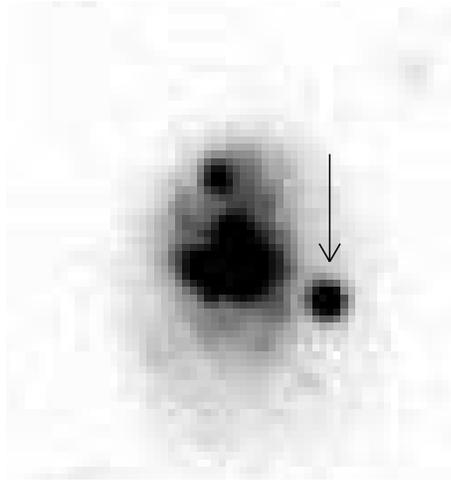} \caption{One of the QSO-galaxy
  pairs where the QSO (SDSS J114719.89+522923.1), which is indicated
  by the arrow, is seen through the galaxy. The impact parameter is 5
  kpc. The redshift of the galaxy and the QSO is $z_G=0.0475$ and
  $z_Q=1.990$.}  \label{fig:image32058}
\end{figure}

Understanding the size and composition of the dust grains in galaxies
provides important astrophysical information. Moreover, it is of
critical importance to correct observations for extinction losses due
to dust, e.g. when measuring cosmological distances with
supernovae. It is often assumed that the reddening properties by dust
in all galaxies follow the average properties of the Milky Way. The
purpose of this work is to examine ways of improving our understanding
of the range of extinction properties. Earlier analyses of dust in
distant galaxies have made use of e.g. multiply imaged QSOs
(e.g. Nadeau et al. \cite{Nadeau}; Falco et
al. \cite{Falco:1999jc}) or galaxy backlitting (e.g. Keel \& White
\cite{Keel2001B}). The method of using multiply imaged QSOs has been
criticised by McGough et al. (\cite{McGough}) since it is implicitly
assumed that one of the images is barely affected by the dust or that
the reddening laws along the two lines-of-sight are very similar. In
addition, a method based on sources found through strong lensing is
only efficient for very massive galaxies, i.e. less suitable for
normal spiral galaxies, especially at low redshifts where the
probability for strong lensing is low. Furthermore, the time
variability of QSOs complicates the interpretation of colours from
different lensed images, as these correspond to different emission
times at the source. In this paper, a method is proposed where the
extinction in galaxies is investigated using observed colours of
single background QSO images shining through or close to foreground
galaxies (as shown in Fig.~\ref{fig:image32058}), taking advantage of
the homogeneous spectral properties of QSOs. With the current
limited data we cannot distinguish between host extinction and
intervening extinction. However, the method will be useful with future
larger samples of QSO-galaxy pairs.

Even though QSOs show small dispersion in their spectral properties, a
population of redder QSOs have been found in studies of QSO colours
(see e.g. Richards et al. \cite{Richards}). These could be explained
invoking extinction by dust, either at the central engine, in the host
galaxy or in a foreground galaxy. There are some indications that the
reddening is dominated by extinction from the host galaxy and only
rarely from an intervening galaxy. Hopkins et al. (\cite{Hopkins})
found that reddened QSOs more often have narrow absorption lines
originating from the QSO redshift than QSOs unaffected by
reddening. No correlation was found between the reddening of QSOs and
the presence of narrow absorption lines originating from smaller
redshifts. In this work, the presence of QSOs affected by intervening
extinction is investigated further by studying pairs of QSOs and
foreground galaxies at small projected distances, comparable to galaxy
sizes. As such systems would also fall into the category of potential
gravitational lensing candidates, we also examine their expected
(achromatic) magnification.

In Sec.~\ref{sec:Reddening}, different parametrisations of dust
attenuation are discussed. In Sec.~\ref{sec:look}, our method for
finding QSOs affected by dust extinction is presented and in
Sec.~\ref{sec:Implementation}, the reddening of a sample of QSOs with
foreground extinction features are investigated. A discussion
follows in Sec.~\ref{sec:Discussion} and a summary of the results in
Sec.~\ref{sec:results}.

Throughout the paper, the cosmological parameters have been assumed to
be $\Omega_M=0.3$, $\Omega_{\Lambda}=0.7$, and $H_0= 70\,{\rm
km/s/Mpc}$.

\section{Reddening by dust}\label{sec:Reddening}

Within the commonly used extinction laws, the specific dust properties
can be quantified by the dimensionless reddening parameter,
the total-to-selective extinction ratio $R_V$, defined by
\begin{equation}
R_V \equiv \frac{A_V}{E(B-V)},
\end{equation}
where $A_V$ is the V-band extinction and $E(B-V)$ is the colour excess,
\begin{equation}
E(B-V) \equiv (B-V)_{\rm obs}-(B-V)_{\rm intr}.
\end{equation}
The observed colour is denoted by $(B-V)_{\rm obs}$ and the intrinsic
colour by $(B-V)_{\rm intr}$. Large grains of dust would produce
(nearly) grey extinction with $R_V \gg 3$. Rayleigh scattering
($A_\lambda \propto \lambda^{-4}$) would lead to a very steep
reddening, $R_V \approx 1.2$.

The typical value found for the reddening parameter in the Milky Way
is $R_V\sim 3.1$, but it has been suggested that it varies between
sight-lines in the interval $2\lesssim R_V \lesssim 6$ (see
e.g. Draine \cite{draine03} and references therein). In other distant
galaxies, values of $R_V$ have been derived in the interval $1.5
\lesssim R_V\lesssim 7.2$ (Falco et al. \cite{Falco:1999jc}), where
the lower values occur in spiral galaxies and the higher values in
ellipticals.

Observations of colours of low-redshift Type Ia supernovae have also
been used to constrain the dust properties of their host galaxies
(e.g. J\"oeveer \cite{joeveer}; Branch \& Tammann
\cite{branch&tammann}; Riess, Press \& Kirshner \cite{Riess}; Reindl
et al. \cite{Reindl}; Guy et al. \cite{guy}). Intriguingly, the
average reddening parameter has been found to be significantly smaller
than in the Milky Way. However, the effect is hard to disentangle from
a possible colour-brightness correlation in supernovae or reddening by
circumstellar dust, possibly significantly different from dust in the
interstellar medium. Thus, an independent investigation using QSO
colours could help to break this degeneracy.

Two parametrisations are often used to describe the extinction law in
our galaxy: Cardelli, Clayton and Mathis (\cite{Cardelli:1989fp}, CCM)
and by Fitzpatrick (\cite{Fitzpatrick:1999}, FTZ). The difference
between these parametrisations are most important for large values of
$R_V$.

For other galaxies, the extinction curves have only been thoroughly
investigated for the Magellanic Clouds, where individual stars can be
resolved. The Magellanic extinction curves are very similar to the
Galactic in the visible and the near-infrared. In the UV, however, the
extinction curve for the Small Magellanic Cloud (SMC) differs
significantly: the curve is almost linear in $\lambda^{-1}$, i.e. the
Galactic type dust bump at 2175 {\AA} is not present. The SMC have
lower dust and metal abundances than the Milky Way and may be more
representative of high redshift galaxies. There is also evidence that
the 2175 {\AA} bump is not present in damped Ly$\alpha$ systems (Pei
et al. \cite{Pei:1999}). Yet, it is often assumed that the Galactic
laws can be applied to dust in other galaxies.


\section{Looking for QSOs affected by intervening extinction}
\label{sec:look}
In order to test the feasibility of finding and using QSOs affected by
dust absorption in a foreground galaxy, the coordinates of the 50\,748
QSOs in the third data release of the Sloan Digital Sky Survey
(Schneider et al. \cite{Schneider:2005}) were cross-correlated with
the positions of the galaxies in the New York University Value-Added
Galaxy Catalog (Blanton et al. \cite{Blanton:2004aa}; Abazajian et
al. \cite{Abazajian:2004aj}). All objects in the galaxy catalogue
where the classification has been disputable have been removed. Among
these were objects within 0.5 arcseconds from a QSO, which often
seemed to be multiple detections of the same source. We also require
that the redshift of the galaxy is determined spectroscopically. Only
QSOs in the redshift range $0.5 < z < 3.0$ were considered, as many of
the QSOs outside this range were found to be less accurately described
by the QSO spectral template made of the HST radio-quiet composite
spectrum (Telfer et al. \cite{Telfer}) combined with the SDSS median
composite spectrum (Vanden Berk et al. \cite{VandenBerk:2001hc}). At
higher redshifts, the flux decrement due to the Ly-$\alpha$ break
makes the bluer bands inefficient.

In the following, a galaxy will be considered as a foreground galaxy
if the distance from the centre of the galaxy to a QSO line-of-sight
(the impact parameter) is less than 30 kpc at the galaxy redshift. The
sample then contains 164 pairs of QSOs and galaxies. The smallest
impact parameter found with respect to the foreground galaxy was 2
kpc. Note that all distances are in units of $(70\,{\rm
km/s/Mpc})/H_0$. Note also that the New York University Value-Added
Galaxy Catalog is very incomplete at $z\gtrsim 0.3$.

For the 164 selected QSOs, the rest-frame $B-V$ colour was estimated
by K-correcting from the observed bands (corrected for Galactic
absorption using the values from SDSS given by the maps of Schlegel et
al. (\cite{schlegel})) using the template QSO spectrum. The QSOs with
foreground galaxies are only slightly redder than the QSOs not
associated with a foreground galaxy (the mean value of $B-V$ differs
by $\sim 0.01$ mag). The two distributions of rest-frame colours were
compared using the Kolmogorov-Smirnov test, yielding a 78\,\%
probability that the difference is only due to statistical
fluctuations. Thus, we conclude that {\em most} of the QSOs of the
selected sample have colours that are indistinguishable from the QSOs
without associated foreground galaxies. Note, however, that this does
not exclude the possibility that both samples contain QSOs reddened by
dust in the host galaxy or in intervening galaxies that may or may not
be resolved. Next, we looked for a possible subset of QSOs in the
selected sample whose colours were best matched with dust extinction
at the redshift of the foreground galaxy. For that purpose we compared
the observed QSO colours with synthetic colours derived from the QSO
spectral template. The template spectrum was redshifted and reddened
as if attenuated by dust at the redshift of the intervening galaxy. A
grid with a wide range of values of $R_V$ and $E(B-V)$ was
investigated using the parametrisation by Fitzpatrick. The expected
colours at the QSO redshift were used to compute the $g$, $r$, $i$,
and $z$ magnitudes from the measured $u$-band magnitude. To find the
values of $R_V$ and $E(B-V)$ that give the best fit for a certain
QSO-galaxy pair, we minimise
\begin{equation}
 \chi^2[R_V,E(B-V)] = (X_{\rm sim}-X_{\rm obs})^T {\cal V}^{-1}(X_{\rm sim}-X_{\rm obs}),
\end{equation}
where $X_{\rm sim}$ and $X_{\rm obs}$ are the simulated and the
observed magnitudes, respectively. ${\cal V}$ is the covariance matrix
calculated from a set of QSOs with similar redshifts ($\Delta z <
0.05$) without any visible foreground galaxies. To avoid a too large
impact from colour outliers, QSOs with colours differing by more than
two standard deviations from the mean were removed in the calculation
of the covariance matrix.

In the process of selecting which of the 164 QSOs that are most
likely to be affected by intervening absorption, the following cuts
were made in order to minimise the number of ``fake'' detections while
still retaining ``good'' candidates with foreground absorption:
\begin{enumerate}
\item The fit of the QSO spectrum with the template without dust 
	extinction had to be worse (higher $\chi^2$) than the fit for
	at least 90\,\% of 10\,000 randomly chosen QSOs without
	visible foreground galaxies.
\item The best fit for dust extinction must be better than the fit for 
	at least 90\,\% of the randomly chosen QSOs affected by a mock
	foreground galaxy.
\end{enumerate}
Two QSO-galaxy pairs survived both cuts. The foreground object was a
spiral galaxy in both of these systems. For the randomly chosen QSOs
without visible intervening galaxies, only 0.2\,\% survived both
cuts. Thus, for the QSO-galaxy sample 0.3 ``fake'' detections are
expected. Using Poisson statistics we thus conclude that the
likelihood for {\em both} events to be false is less than 4\,\%
($P(n)=0.74,0.22,0.03$, for $n=0,1,2$ respectively). However, we 
cannot exclude the possibility that the two selected quasars 
are colour outliers in the poorly understood tail of the 
colour distribution.

\begin{figure}
  \centering
  \includegraphics[bb= 125 125 300 280, clip=true,width=6cm]{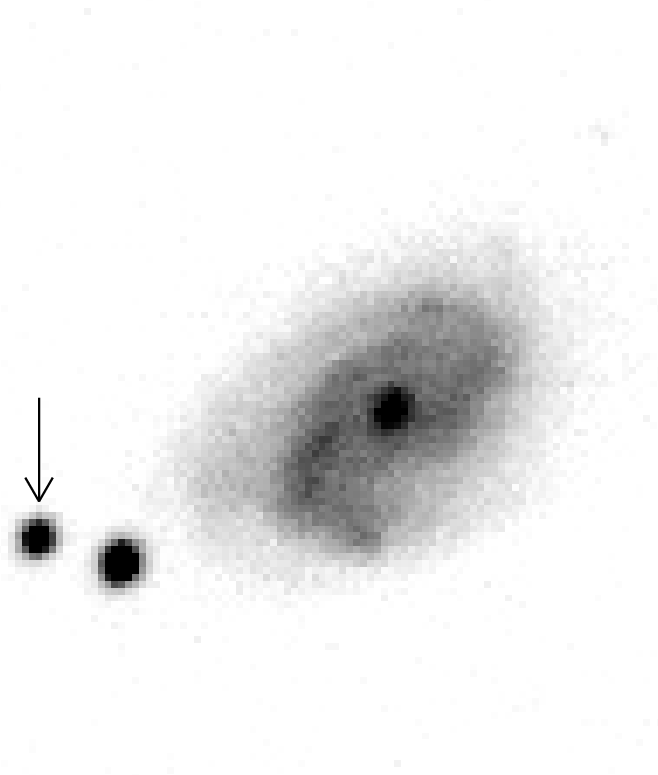}
  \includegraphics[bb= 120 125 320 260,clip=true,width=6cm]{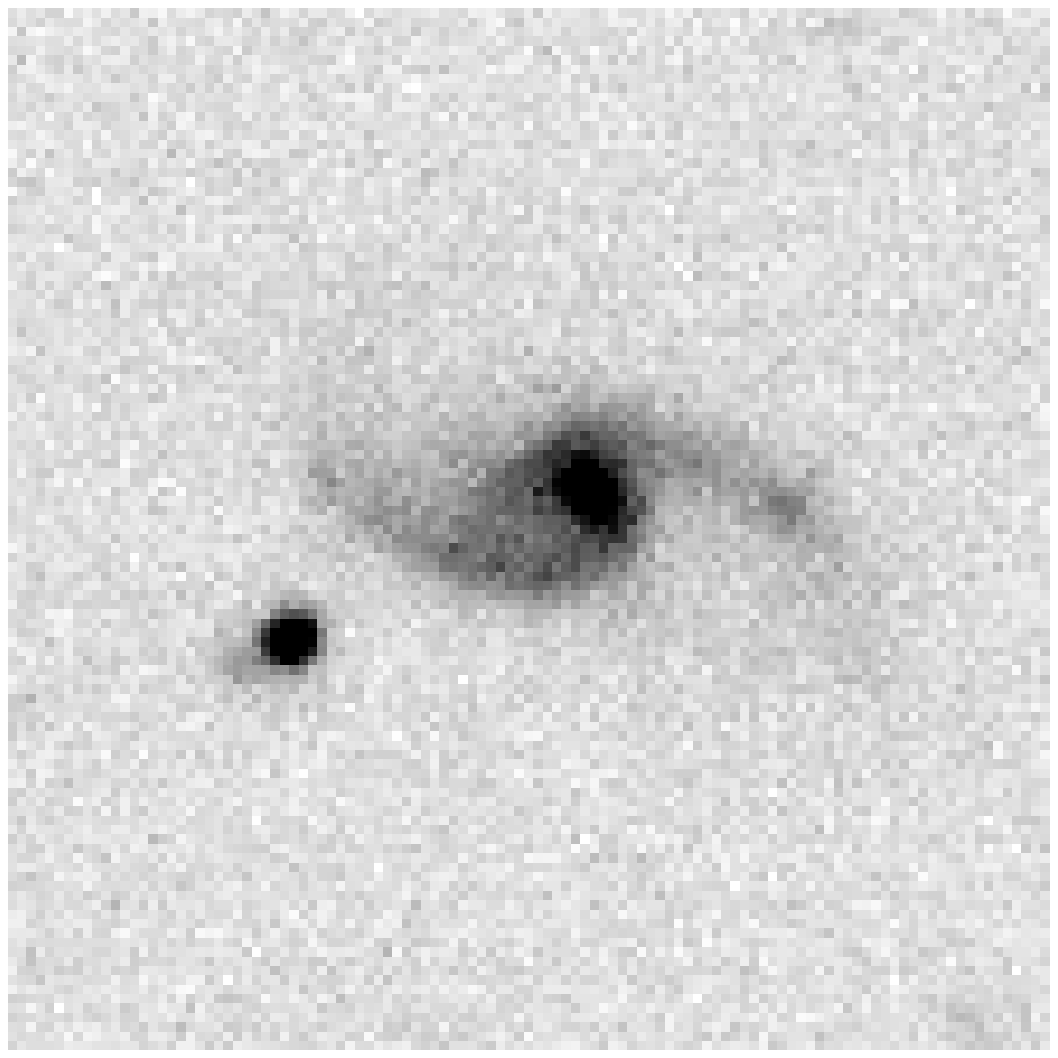}
  \caption[]{The two QSO-galaxy pairs that survived the cuts,
  SDSS J131058.13+010822.2, and SDSS J084957.97+510829.0 from the
  top. An arrow points out the QSO in the case where it is not
  clear which object it is.}
  \label{fig:image1379}
\end{figure}

The low number of pairs surviving both cuts consolidates our earlier
conclusion that the bulk of the sample with 164 QSOs with foreground
galaxies are not significantly different from the whole sample. This
is also confirmed by a visual inspection of all QSO-galaxy pairs
showing that, due to the relatively large impact parameters, only two
of the QSOs are clearly behind any visible part of the foreground
galaxy. These two, however, did not make it through the cuts. Both of
the QSOs that survived the cuts have an impact parameter close to, but
clearly outside the visible parts of the foreground galaxy see
Fig.~\ref{fig:image1379}. In both pairs, the surface brightness of the
galaxy at the position of the QSO is of the order 0.1 \% of the
central galaxy surface brightness. This fact puts limitations to
the possible intervening extinction, unless there are significant
amounts of dust outside the optical disc of spiral galaxies. In the
following, these two pairs will be used to demonstrate our method, even
though it is possible that the derived extinction is not only due to
foreground extinction, but could also be due to host extinction.

\begin{figure}
   \includegraphics[width=8cm]{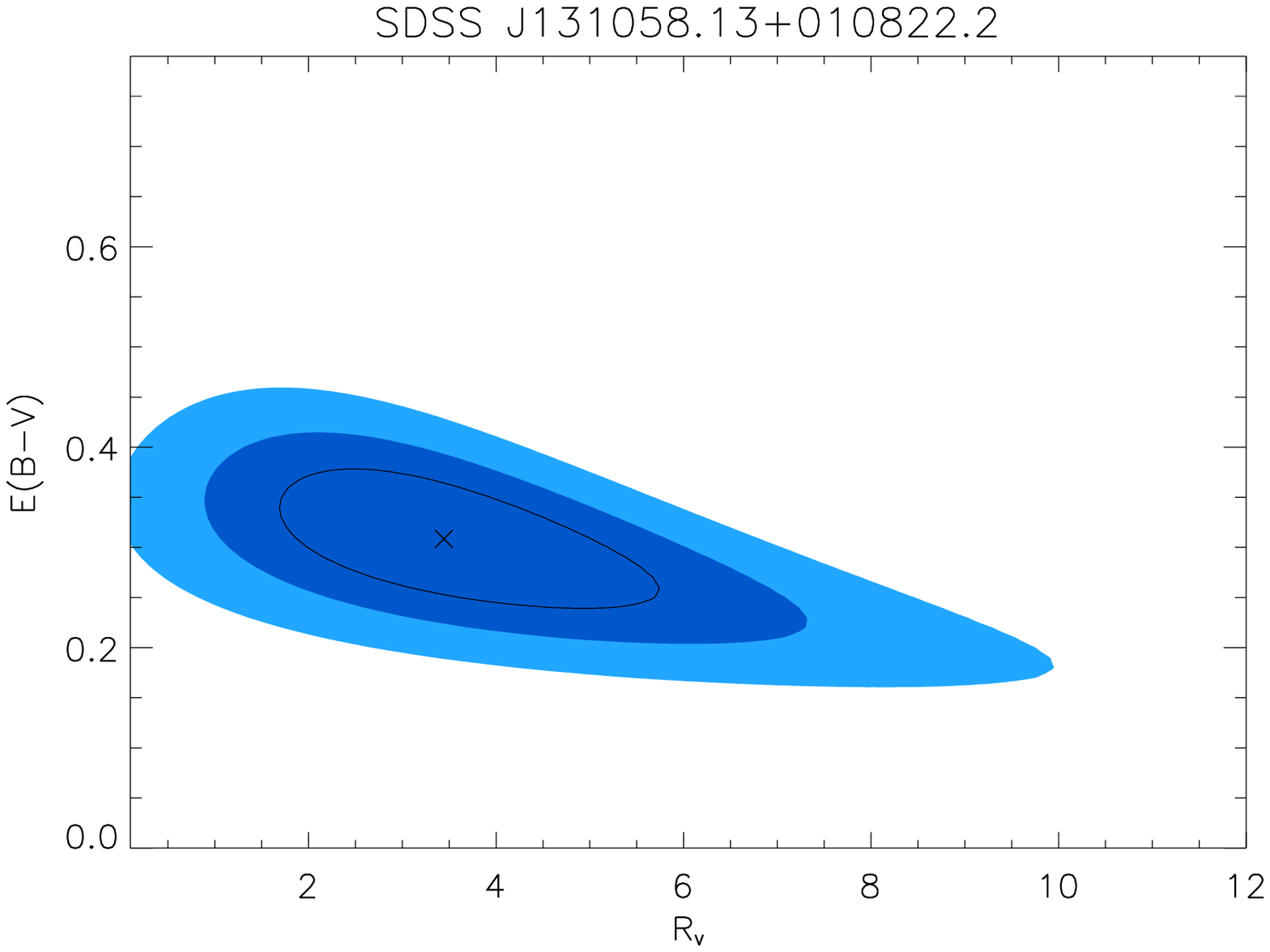}
   \includegraphics[width=8cm]{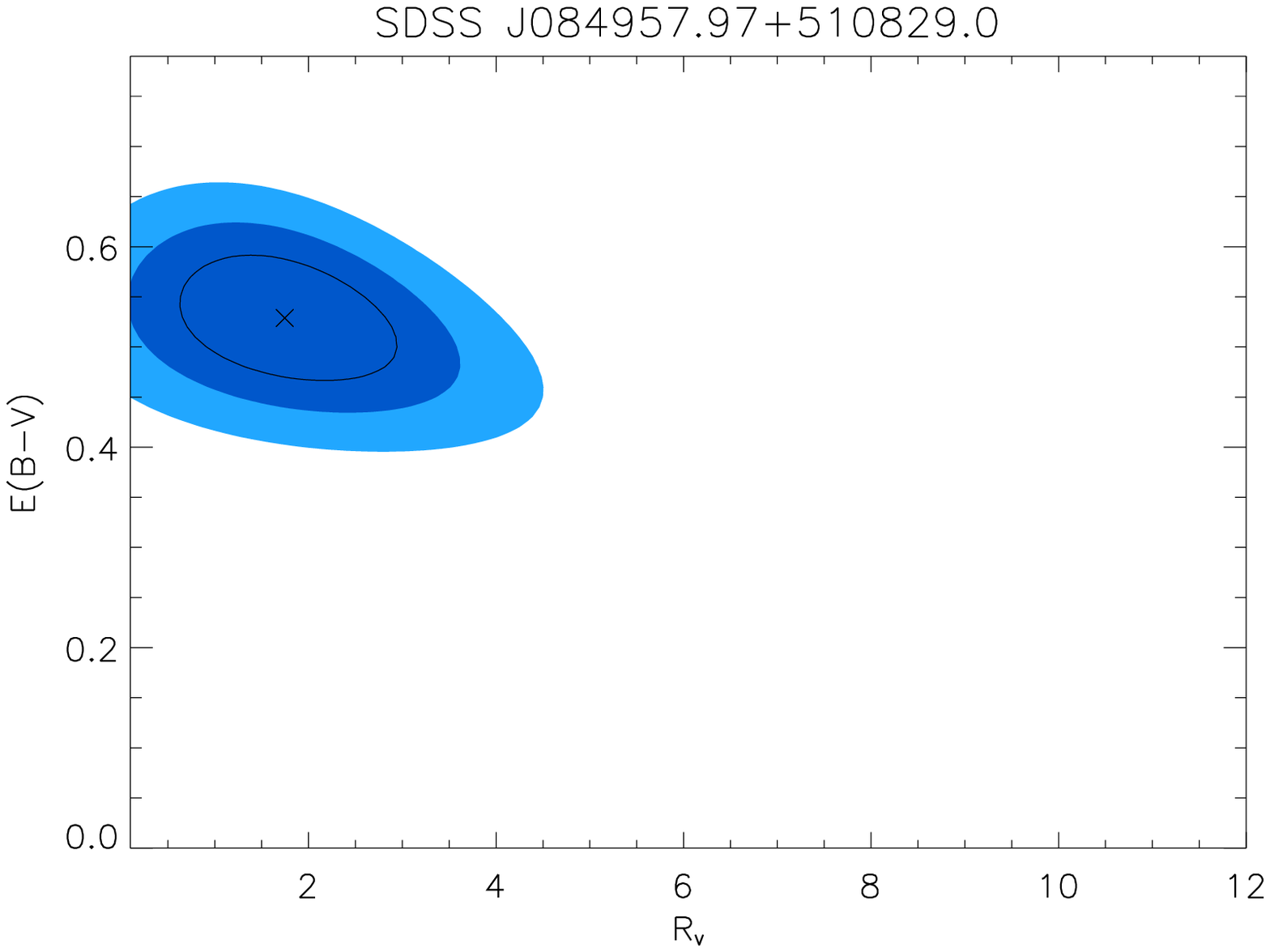}
   \caption{Confidence levels for the two low-redshift galaxies
   corresponding to 1 $\sigma$ (black line), 68\,\% (dark blue region)
   and 90\,\% (pale blue) as defined by the $\chi^2$ test.}
   \label{fig:conf2}
\end{figure}

Information about the two remaining QSO-galaxy pairs, as well as the
fitted dust parameters are listed in Table~\ref{table:info}.
\begin{table*}
\caption{Two colour selected QSO-galaxy pairs. The first two columns
contain data about the QSO and the following four about the
galaxy. The positions are given in J2000 and $R_P$ is the petrosian
radius in the R band from SDSS. The impact parameters are then given
both as the angle and the distance at the galaxy redshift. The last
two columns contain the best fit values of $R_V$ and the one $\sigma$
uncertainty for two different dust parametrisations, Cardelli, Clayton
and Mathis (\cite{Cardelli:1989fp}, ccm) and Fitzpatrick
(\cite{Fitzpatrick:1999}, ftz).}
\label{table:info}
\centering
\begin{tabular} {c c | c c c c | c c | c c}
\hline \hline
DR3 Object Designation & $z_Q$ & ${\rm RA}_G$ & ${\rm DEC}_G$ & $z_G$ & $R_P$ & Impact ('') & Impact 
(kpc) & $R_{V,ccm}$ & $R_{V,ftz}$ \\
\hline
J131058.13+010822.2 & 1.389 & 197.744 & 1.145 & 0.0357 & 12'' & 22'' & 15 & $3.4_{-1.7}^{+2.3}$ & $3.4_{-1.0}^{+1.3}$\\
J084957.97+510829.0 & 0.584 & 132.490 & 51.145 & 0.0734 & 10'' & 14'' & 20 & $1.7_{-1.0}^{+1.2}$ & $2.2_{-0.7}^{+0.7}$\\
\hline \\
\end{tabular}
\end{table*}
The best fit values of $R_V$ are listed both for the parametrisation
by Fitzpatrick and the CCM parametrisation together with the error
interval under the assumption that the extinction law is valid for all
positive $R_V$ values. The $\chi^2$ is generally somewhat better for
the Fitzpatrick parametrisation. In both cases the allowed values for
$R_V$ are consistent with the Galactic average ($R_V\sim3.1$), as
shown in Fig.\ref{fig:conf2} for the CCM parametrisation. The
parametrisation by Fitzpatrick gives smaller contours (not shown).
We have also modeled the quasar colours with extinction by SMC type
dust but the fit is worse than for the CCM and FTZ parametrisations.

As noted in Hopkins et al. (\cite{Hopkins}), the reddening in the
selected systems could originate in the host galaxy, but such
configurations yield worse fits in both cases. Of course, the
reddening could also be a combination of both foreground and host
extinction.


\section{Reddening in QSOs shining through galaxies}
\label{sec:Implementation}

Since the New York University Value-Added Galaxy Catalog is very
incomplete at $z\gtrsim 0.3$, we have searched the literature for
QSO-galaxy pairs with higher galaxy redshifts. Wang et
al. (\cite{wang}) have found by a visual inspection three QSO spectra
in the Sloan Digital Sky Survey (Schneider et
al. \cite{Schneider:2005}) with a 2175 {\AA} absorption feature at a
lower redshift than for the QSO. These QSOs are therefore most likely
affected by dust in an intervening galaxy. The redshift of the
intervening system was determined using multiple narrow absorption
lines. In addition, Pindor et al. (\cite{pindor}) and Johnston et
al. (\cite{johnston}) have found doubly imaged QSOs where the redshift
of the lensing galaxy has been determined from absorption features.

To explore the possibility of dust absorption in these individual QSO
images, the observed QSO colours were compared with the simulated
colours for different values of $R_V$ and $E(B-V)$, in the same way as
described in Sec.~\ref{sec:look}. Note that we have not used the
differential method that is traditionally used for determining dust
extinction using multiply images mentioned in
Sec.~\ref{sec:intro}. The best fit values for $R_V$ allowing only
positive $R_V$ are given in Table~\ref{table:info2} for the two
different parametrisations. The two last entries in the table
correspond to fits using two different images of the same QSO. In the
following, we will use the mean of these two values of $R_V$. None of
the pairs in the table would have survived the cuts in
Sec.~\ref{sec:look}. It is intriguing that the best-fit values of
$R_V$ appear to be significantly lower than for Galactic dust.
\begin{table*}
\caption{Information about QSO-galaxy pairs from the literature. The
first two columns contain data about the QSO. The third column
contains the galaxy redshift and the following four columns contain
our best fit values of $R_V$ and $E(B-V)$ for two different dust
parametrisations and the one $\sigma$ uncertainty. The last column
contains the reference from where the pair was found.}
\label{table:info2}
\centering
\begin{tabular} {c c | c | c c c c | c}
\hline \hline
Object Designation & $z_Q$ & $z_G$ & $R_{V,ccm}$ & $E(B-V)_{ccm}$ & $R_{V,ftz}$  & $E(B-V)_{ftz}$ & Source \\
\hline
J144612.98+035154.4 & 1.945 & 1.512 & $0.0_{-0.0}^{+0.7}$& $0.30_{-0.03}^{+0.03}$& $1.3_{-0.3}^{+0.3}$ & $0.25_{-0.06}^{+0.07}$ &  Wang et al. (\cite{wang})\\
J145907.19+002401.2 & 3.011 & 1.395 & $0.0_{-0.0}^{+2.5}$ & $0.13_{-0.02}^{+0.07}$& $1.8_{-1.4}^{+1.7}$ &  $0.16_{-0.13}^{+0.18}$ & Wang et al. (\cite{wang}) \\
J012147.73+002718.7 & 2.225& 1.388 & $3.3_{-3.2}^{+1.2}$ & $0.24_{-0.14}^{+0.18}$ &$1.8_{-1.3}^{+1.2}$ & $0.12_{-0.09}^{+0.11}$ & Wang et al. (\cite{wang}) \\
J090335.15+502820.2 & 3.584 & 0.388 & $0.0_{-0.0}^{+0.7}$ & $0.61_{-0.03}^{+0.04}$ & $0.9_{-0.2}^{+0.3}$ & $0.51_{-0.09}^{+0.07}$ & Johnston et al. (\cite{johnston})\\
J115517.34+634622.0 & 2.888 & 0.176 & $0.8_{-0.7}^{+1.7}$& $0.52_{-0.09}^{+0.09}$ &$1.3_{-0.6}^{+1.0}$ & $0.51_{-0.11}^{+0.09}$  & Pindor et al. (\cite{pindor})\\
J115517.08+634621.5 & 2.888 & 0.176 & $0.0_{-0.0}^{+1.3}$& $0.33_{-0.08}^{+0.09}$ &$0.6_{-0.4}^{+1.2}$ &  $0.26_{-0.13}^{+0.11}$ & Pindor et al. (\cite{pindor})\\
\hline \\
\end{tabular}
\end{table*}


\section{Discussion}
\label{sec:Discussion}

\begin{figure}
  \centering
  \includegraphics[width=8cm]{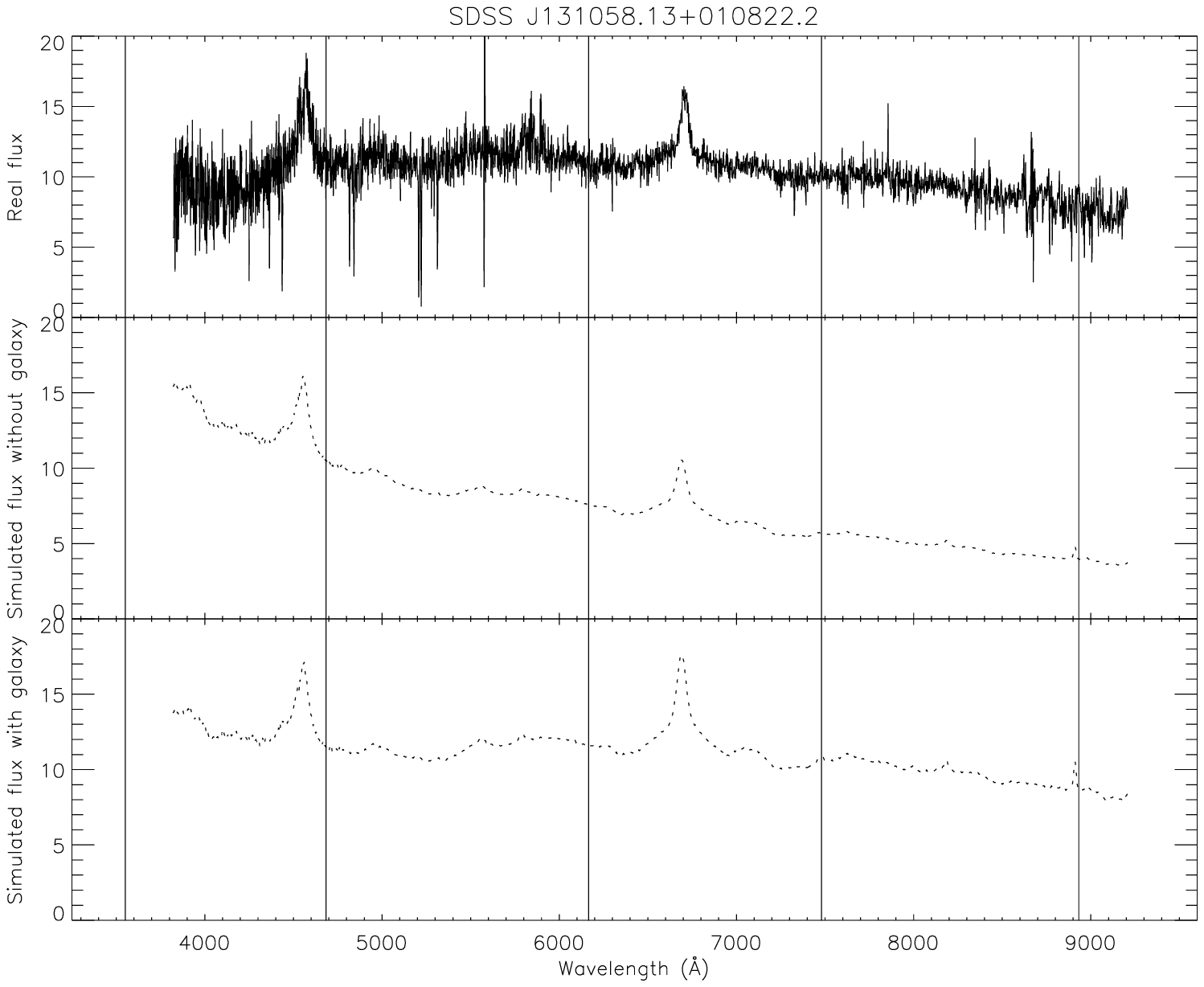}
  \includegraphics[width=8cm]{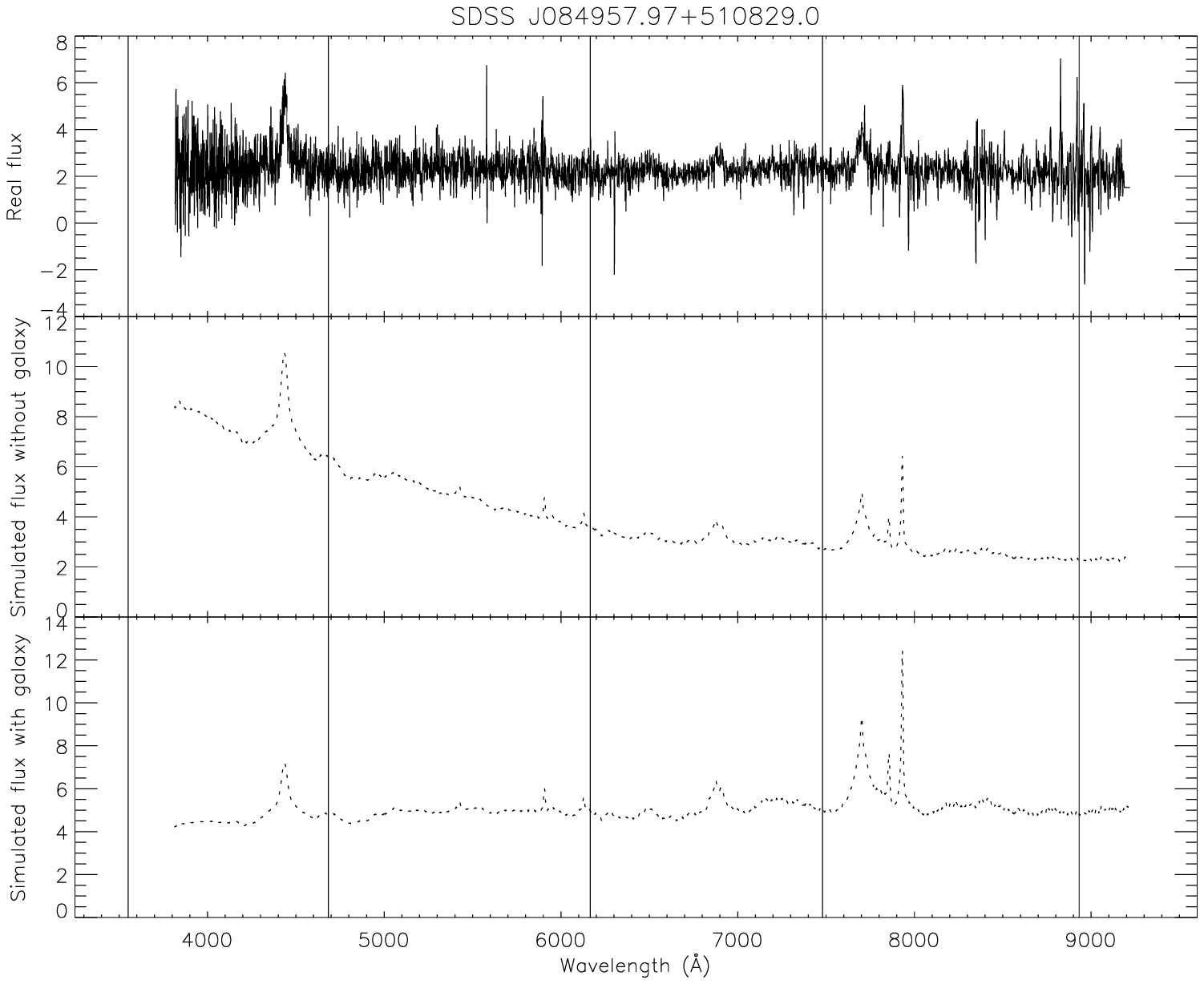}
  \caption{The real spectrum, the redshifted template spectrum and the
  template spectrum after being redshifted and reddened due to the
  best fit dust with the Fitzpatrick parametrisation for the two
  QSO-galaxy pairs. The vertical lines show the average wavelengths of
  the broadband filters.}
\label{fig:spectra}
\end{figure}

Our QSO colour analysis makes only use of the SDSS QSO
photometry. However, a visual comparison was also made between the
candidate spectra and the redshifted template spectrum with and
without reddening following the best fit values of $E(B-V)$ and
$R_V$. We find that the two candidate spectra are consistent with the
fit based on the photometric data. These comparisons are shown in
Fig.~\ref{fig:spectra}. However, the spectroscopic comparison could
not rule out host extinction as opposed to extinction in a foreground
galaxy.

\begin{figure}
  \centering \includegraphics[width=8cm]{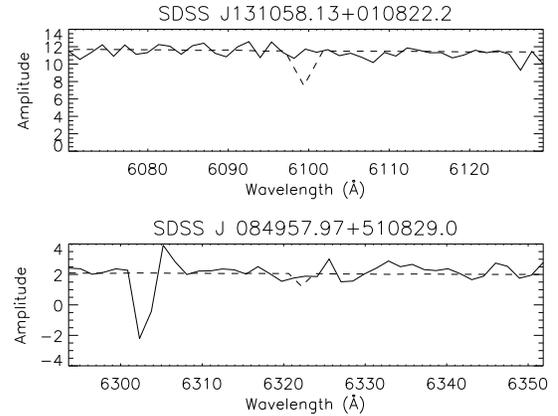} \caption{The
  SDSS spectrum of the two QSOs with a low-redshift galaxy (solid
  lines) and the redshifted template spectrum with a dip corresponding
  to the absorption from Na I in the intervening galaxy if $E(B-V)$ is
  given by the best fit value.} \label{fig:zoomspectra}
\end{figure}

\begin{figure}
  \centering \includegraphics[width=7cm]{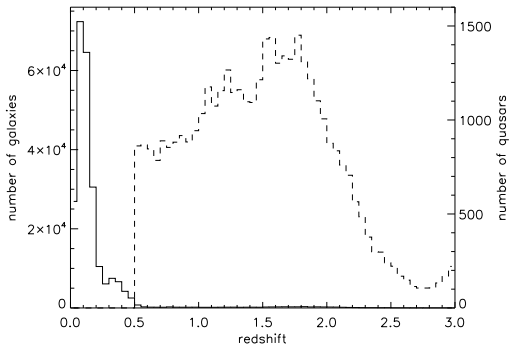}
  \caption{The redshift distribution of the galaxies in the NYU VAGC
  (solid lines) and of the QSOs in DR3 of the SDSS with $0.5<z<3.0$
  (dotted lines).}  \label{fig:zdistr}
\end{figure}

We have also looked for narrow absorption lines in the QSO
spectra. The redshift of the line would tell whether the absorption
occurs in the host galaxy or in an intervening galaxy. Munari and
Zwitter (\cite{Munari}) have found a relation between the equivalent
width of Na I (5890.0 \AA) and K I (7699.0 \AA) and $E(B-V)$. This
could provide a second test for $E(B-V)$ if the lines are detected.
Unfortunately, the signal-to-noise of the SDSS spectra is not good
enough to put limits on $E(B-V)$, as shown in
Fig.~\ref{fig:zoomspectra}. Thus, this is not testable with the public
SDSS spectroscopic data, but would require follow-up observations with
higher spectral resolution and better S/N-ratio.
The detection of such absorption lines in QSO-galaxy pairs would
provide evidence for the existence of dust outside the optical disc of
spiral galaxies, indicative of strong expulsion from dust producing
regions (Ferrara et al. \cite{Ferrara1990}; Aguirre
\cite{Aguirre1999}). Without such confirmation, the fact that we only
have two candidates for foreground galaxy extinction that both
correspond to positions at large optical radii implies that we need to
be careful when interpreting the data. Since the galaxy catalogue is
not as deep as the QSO catalogue (see Fig.~\ref{fig:zdistr}), we
expect to suffer from both incompleteness in our sample of QSO-galaxy
pairs and contamination in our reference sample.
With a deeper galaxy survey more QSO-galaxy pairs would be found, also
with smaller impact parameters. Also, a deeper survey would reduce the
uncertainty in the calculation of the covariance matrix due unresolved
foreground objects in the reference QSO sample. Forthcoming wide
field surveys, e.g. the Large Synoptic Survey Telescope
(LSST)\footnote{\tt http://www.lsst.org}, could potentially provide
over a thousand QSO-galaxy pairs where the dust properties in the
foreground galaxy could be successfully measured as described in
Sec.~\ref{sec:Implementation}.

Systems consisting of a high redshift QSO with a foreground galaxy
close to the line-of-sight are not only interesting from a dust
absorption point of view; it is also a classical set up for a
gravitational lens system. The light from the background QSO will
become deflected when passing through the gravitational field of the
foreground galaxy causing images to be magnified, distorted and - in
rare cases - multiply imaged. In order to constrain the magnitude of
the lensing effect, we model the mass distribution of the foreground
galaxies as singular isothermal spheres (SIS) with density profiles
\begin{equation}
\rho_{\rm SIS}(r)=\frac{\sigma^2}{2 \pi r^2}.
\end{equation}
The model is characterised by the line-of-sight velocity dispersion
$\sigma$, which can be estimated from the galaxy luminosity via the
Faber--Jackson relation (elliptical and lenticular galaxies) or the
Tully--Fisher relation (spiral galaxies), see e.g. Gunnarsson et
al. (\cite{gunnar05}) and references therein.

For each foreground galaxy, the absolute B-band magnitude, $M_B$, is
estimated using the publicly available code {\tt kcorrect v3\_2}
(Blanton et al. \cite{Blanton:2003}). The galaxy type is determined
from the observed galaxy colours and spectra in combination with
visual inspections of the morphologies.

For the sample of 164 QSO-galaxy pairs with impact parameter less than
30 kpc at the galaxy redshift, magnifications are generally small
($\mu \lesssim 1.1$). The only case where our calculations gives a
magnification larger than 1.1 is the pair shown in
Fig.~\ref{fig:image32058} which has one of the smallest impact
parameter of our pairs. Provided the lensing galaxy is an elliptical,
the magnification factor for this QSO is $\mu\sim 1.3$. The system has
an unusual morphology and we cannot exclude the possibility that a
more careful lensing analysis might give more interesting results.
However, such detailed investigations are outside the scope of this
paper.


\section{Summary and conclusions}
\label{sec:results}

We have investigated the possibility to detect the reddening of QSOs
by foreground galaxies and to measure the dust extinction properties.
Cross-correlating coordinates of the SDSS quasar catalogue with the
New York University Value-Added Galaxy Catalog we find 164 QSO-galaxy
pairs with impact parameters within 30 kpc at the redshift of the
foreground galaxy. With additional colour cuts we identified two
low-$z$ systems ($z<0.08$) involving spiral galaxies where the QSO
colours are particularly indicative of reddening in the intervening
galaxy. Both QSOs have very large impact parameters in terms of the
the optical radii of the foreground galaxies and are thus indicative
of large dust column densities outside the optical disc of spiral
galaxies. With this assumption, and using redshifted reddened QSO
templates we fit for $R_V$ and $E(B-V)$ using the measured colours and
redshifts of the quasar and galaxy and find that Galactic type dust
provides a good fit to the data. Such an interpretation needs to be
confirmed by e.g.~spectroscopic means. We emphasize that we cannot
rule out other explanations for the red colour of these two quasars,
e.g.  intrinsic redder colour, extinction in the host galaxy or in an
unresolved foreground galaxy.

We also explored the extinction properties in five QSOs in the
literature, where the foreground galaxies have redshifts in the range
$z=0.15-1.5$. A wide range of preferred values of $R_V$ was
found. This is an indication that the dust properties in other
galaxies may be different from the Milky Way.

\begin{figure}
  \centering \includegraphics[width=8cm]{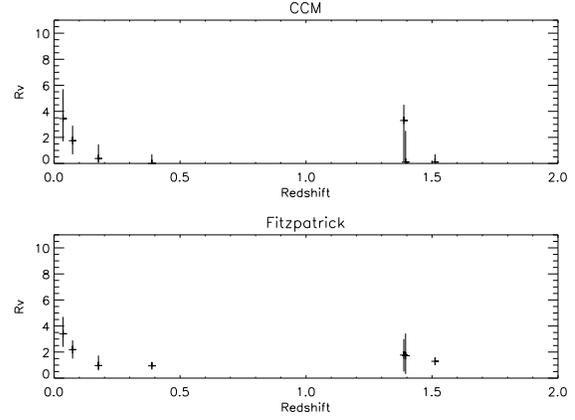} \caption{The
  best fit values of $R_V$ ordered after galaxy redshift for the CCM
  and Fitzpatrick parametrisations.}
\label{fig:rvdistr}
\end{figure}

We note with interest that the dust in galaxies with high redshifts in
our study were better fitted by lower values of $R_V$ than the average
Milky Way value (See Fig.~\ref{fig:rvdistr}).
If this effect is confirmed with a larger sample, and selection
effects can be ruled out, it would be indicative of an evolution of
the dust properties with redshift, which would have an impact on
e.g. the use of extinction corrected supernovae as distance
indicators. It would not be too surprising if such an evolution in the
reddening law exists since e.g. metallicity and elemental abundance
ratios are known to change with redshift and these are expected to
affect the dust properties. However, it is possible that the
differences between our two sets are mainly due to selection effects
since the few systems studied have been chosen with different methods:
at high-$z$ through detection of absorption features and at low-$z$ by
selecting reddened QSOs at small impact parameters with respect to
resolved foreground galaxies.


\begin{acknowledgements}

The authors would like to thank Vallery Stanishev and the anonymous
referee for many useful comments. We would also like to thank the
G\"{o}ran Gustafsson Foundation for financial support. EM would like
to thank the Royal Swedish Academy of Sciences for financial support.
AG is a Royal Swedish Academy Research Fellow supported by grants from
the Swedish Research Council, the Knut and Alice Wallenberg
Foundation.

Funding for the Sloan Digital Sky Survey (SDSS) has been provided by
the Alfred P. Sloan Foundation, the Participating Institutions, the
National Aeronautics and Space Administration, the National Science
Foundation, the U.S. Department of Energy, the Japanese
Monbukagakusho, and the Max Planck Society. The SDSS Web site is
http://www.sdss.org/. The SDSS is managed by the Astrophysical
Research Consortium (ARC) for the Participating Institutions. The
Participating Institutions are The University of Chicago, Fermilab,
the Institute for Advanced Study, the Japan Participation Group, The
Johns Hopkins University, Los Alamos National Laboratory, the
Max-Planck-Institute for Astronomy (MPIA), the Max-Planck-Institute
for Astrophysics (MPA), New Mexico State University, University of
Pittsburgh, Princeton University, the United States Naval Observatory,
and the University of Washington.

\end{acknowledgements}


\begin{thebibliography}{}

\bibitem[2004]{Abazajian:2004aj} Abazajian, K., Adelman-McCarthy, J.~K., Ag\"{u}eros, M.~A., et al. 2004, AJ, 128, 502

\bibitem[1999]{Aguirre1999} Aguirre, A.~N. 1999, ApJ, 512, L19

\bibitem[2003]{Blanton:2003} Blanton, M.~R., Brinkmann, J., Csabai, I., et al. 2003, AJ, 125, 2348

\bibitem[2005]{Blanton:2004aa} Blanton, M.~R., Eisenstein, D., Hogg, D.~W., Schlegel, D.~J., \& Brinkmann, J. 2005, ApJ, 629, 143

\bibitem[1992]{branch&tammann} Branch, D., \& Tammann, G.A. 1992, ARA\&A, 30, 359

\bibitem[1989]{Cardelli:1989fp} Cardelli, J.~A., Clayton, G.~C., \& Mathis, J. S. 1989, ApJ, 345, 245-256

\bibitem[2003]{draine03} Draine, B.~T. 2003, ARA\&A, 41, 241

\bibitem[1999]{Falco:1999jc} Falco, E.~E., Impey, C.~D., Kochanek, C.~S., et al. 1999, ApJ, 523, 617

\bibitem[1990]{Ferrara1990} Ferrara, A., Aiello, S., Ferrini, F., \& Barsella, B. 1990, A\&A, 240, 259

\bibitem[1999]{Fitzpatrick:1999} Fitzpatrick, E.~L. 1999, PASP, 111, 63-75

\bibitem[2005]{gunnar05} Gunnarsson, C., Dahl\'en, T., Goobar, A., J\"onsson, J., \& M\"{o}rtsell, E. 2005, ApJ {\em in press}, astro-ph/0506764

\bibitem[2005]{guy} Guy, J., Astier, P., Nobili, S., Regnault, N., \& Pain, R. 2005, A\&A, 443, 781

\bibitem[2004]{Hopkins} Hopkins, P.~F., Strauss, M.~A.,Hall,
P.~B., et al. 2004, AJ, 128, 1112

\bibitem[2003]{johnston} Johnston, D.~E., Richards, G.~T.,
Frieman, J.~A., et al. 2003, AJ, 126, 2281

\bibitem[1983]{joeveer} J\"oeveer, M. 1983, Afz, 18, 328

\bibitem[2001]{Keel2001B} Keel, W.~C., \& White, R.~E. III 2001, AJ, 122, 1396

\bibitem[2005]{McGough} McGough, C., Clayton, G.~C., Gordon,
K.~D., \& Wolff, M.~J. 2005, ApJ, 624, 118-123

\bibitem[1997]{Munari} Munari, U., \& Zwitter, T. 1997, A\&A, 318, 269-274

\bibitem[1991]{Nadeau} Nadeau, D., Yee, H.~K.~C., Forrest, W.~J., et al. 1991, ApJ, 376, 430

\bibitem[1999]{Pei:1999} Pei, Y.~C., Fall, S.~M., \& Hauser, M.~G. 
	1999, ApJ, 522, 604

\bibitem[2003]{pindor} Pindor, B., Eisenstein, D.~J., Inada, N.,
et al. 2004, AJ, 127, 1318

\bibitem[2005]{Reindl} Reindl, B., Tammann, G.~A., Sandage, A.,
\& Saha, A. 2005, ApJ, 624, 532

\bibitem[2001]{Richards} Richards, G.~T., Fan, X., Schneider, D.~P., et al. 2001, AJ, 121, 2308

\bibitem[1996]{Riess} Riess, A.~G., Press, W., \& Kirshner, R.~P. 1996, ApJ, 473,588

\bibitem[1998]{schlegel} Schlegel, D.~J., Finkbeiner, D.~P., \& Davis, M. 1998, ApJ, 500, 525

\bibitem[2005]{Schneider:2005} Schneider, D.~P., Hall, P.~B., Richards, G.~T., et al. 2005, AJ, 130, 367

\bibitem[2002]{Telfer} Telfer, R.~C., Zheng, W., Kriss, G.~A., \&
Davidsen, A.~F. 2002, ApJ, 565, 773

\bibitem[2001]{VandenBerk:2001hc} Vanden Berk, D. E., Richards,
G.~T., Bauer, A., et al. 2001, AJ, 122, 549-564

\bibitem[2004]{wang} Wang, J., Hall, P.~B., Ge, J., Li, A., \&
Schneider, D.~P. 2004, ApJ, 609, 589




\end{thebibliography}
\end{document}